\documentclass[12pt, a4paper]{article}
\usepackage{amsmath,amsfonts}

\def\halft{{\textstyle\frac{1}{2}}}

\begin{document}

\numberwithin{equation}{section}

\title{Inhomogeneous ``longitudinal'' plane waves \\
       in a deformed elastic material.}
\author{Michel Destrade \& Michael Hayes}
\date{}
\maketitle
%

\begin{abstract}

By definition, a homogeneous isotropic compressible Hadamard material 
has the property that an infinitesimal longitudinal homogeneous plane 
wave may propagate in every direction when the material is
maintained in a state of arbitrary finite static homogeneous
deformation. 
Here, as regards the wave, `homogeneous' means that the direction of
propagation of the wave is parallel to the direction of eventual
attenuation; and `longitudinal' means that the wave is linearly
polarized in a direction parallel to the direction of propagation.
In other words, the displacement is of the form
$\mathbf{u} = \mathbf{n} \cos k(\mathbf{n \cdot x} - ct)$,
where $\mathbf{n}$ is a real vector.

It is seen that the Hadamard material is the most general
one for which a `longitudinal' \textit{inhomogeneous} plane wave may
also propagate in any direction of a predeformed body. 
Here, `inhomogeneous' means that the wave is attenuated, in a direction
distinct from the direction of propagation; and ``longitudinal'' means
that the wave is elliptically polarized in the plane containing these
two directions, and that the ellipse of polarization is similar and
similarly situated to the ellipse for which the real and imaginary
parts of the complex wave vector are conjugate semi-diameters.
In other words, the displacement is of the form
$\mathbf{u}
 = \Re \{\mathbf{S} \exp i \omega (\mathbf{S \cdot x} - ct) \}$,
where $\mathbf{S}$ is a complex vector (or bivector).

Then a Generalized Hadamard material is introduced. 
It is the most general homogeneous isotropic compressible material 
which allows the propagation of infinitesimal 
``longitudinal'' inhomogeneous plane circularly polarized waves 
for all choices of the isotropic directional bivector.

Finally, the most general forms of response functions are found
for homogeneously deformed isotropic elastic materials in which 
``longitudinal'' inhomogeneous plane waves may propagate with a 
circular polarization in each of the two planes of central circular 
section of the $\mathbb{B}^n$-ellipsoid, where $\mathbb{B}$ is the 
left Cauchy-Green strain tensor corresponding to the primary pure 
homogeneous deformation.

\end{abstract}

\section{Introduction.}

Infinitesimal longitudinal homogeneous plane waves play a special role 
in classical linearized theory.
For such waves the amplitude is parallel to the direction of 
propagation $\mathbf{n}$ so that all the particles oscillate along the 
direction $\mathbf{n}$. 
Such waves may propagate in every direction in an isotropic 
compressible elastic body.
This is no longer the case for propagation in an elastic anisotropic 
crystal. 
Possible directions of propagation of longitudinal homogeneous plane 
waves, called `specific directions' by Borgnis \cite{Borg55}, may be 
as few as three in an elastic anisotropic crystal.
Of course, by assuming certain restrictions on the elastic constants, 
it is possible, as Hadamard \cite{Hada03} did, to have a special model 
anisotropic material in which longitudinal waves are possible in every 
direction.

A similar situation holds for infinitesimal motions of an isotropic 
homogeneous elastic material held in an (arbitrary) state of static 
(finite) homogeneous deformation -- specific directions are limited in 
number. 
However, it is possible to place conditions on the basic strain energy 
density, or on the response coefficients defining the material, such 
that every direction is specific for arbitrary (finite) homogeneous 
deformation. 
The resulting model was called a `Hadamard material' by John 
\cite{John66}.

Here we consider the propagation of infinitesimal 
\textit{inhomogeneous} plane waves in an isotropic homogeneous 
compressible elastic body held in an arbitrary state of finite static 
homogeneous deformation. 
Such waves may be described in terms of bivectors 
-- complex vectors -- the amplitude bivector $\mathbf{A}$ and the 
slowness bivector $\mathbf{S}$, which is written 
$\mathbf{S} = N \mathbf{C}$, where the directional bivector 
$\mathbf{C}$ is written 
$\mathbf{C} = m \mathbf{\hat{m}} + i \mathbf{\hat{n}}$ 
($\mathbf{\hat{m} \cdot \hat{n}} =0$, $m \ge 1$,  
$|\mathbf{\hat{m}}| = |\mathbf{\hat{n}}| = 1$). 
Once the directional bivector $\mathbf{C}$ is prescribed, the slowness 
$\mathbf{S}$ and the amplitude $\mathbf{A}$ are determined from the 
equations of motion. 
Prescribing $\mathbf{C}$ is equivalent to prescribing an ellipse with 
principal semi-axes  $ m \mathbf{\hat{m}}$ and $\mathbf{\hat{n}}$; 
this directional ellipse for inhomogeneous plane waves corresponds to 
the direction of propagation $\mathbf{n}$ for homogeneous plane waves 
\cite{Haye84}. 
We borrow the adjective `longitudinal' to describe an inhomogeneous 
plane wave for which $\mathbf{A}$ and $\mathbf{S}$ (and $\mathbf{C}$) 
are parallel: $\mathbf{A} \times \mathbf{S} = \mathbf{0}$. 
What this means is that \cite{BoHa93} the ellipses of $\mathbf{A}$ and 
of $\mathbf{S}$ (and of $\mathbf{C}$) are all parallel, similar 
(same aspect ratio) and similarly situated (parallel major axes and 
parallel minor axes). 
We determine the most general form of strain-energy density so that 
such waves may propagate for all choices of the directional bivector 
$\mathbf{C}$. 
It turns out to be a Hadamard material.

Then we consider the possibility of having `circularly polarized 
longitudinal' waves. 
For such waves both $\mathbf{C}$ and $\mathbf{A}$ are isotropic and 
coplanar: the corresponding ellipses are coplanar circles. 
Such waves are a subclass of longitudinal waves. 
The corresponding class of materials for which circularly polarized 
longitudinal waves are possible is called a `Generalized Hadamard 
material'. 
Clearly they include Hadamard materials. 

Next we obtain constitutive equations for materials which allow the 
propagation of two special infinitesimal longitudinal circularly 
polarized inhomogeneous plane waves for all choices of the basic  
static homogeneous deformation.
The circles of polarization are to lie in the planes of central 
circular section of the $\mathbb{B}^n$-ellipsoid: 
$\mathbf{x \cdot}\mathbb{B}^n \mathbf{x} =1$, where $\mathbb{B}$ is 
the left Cauchy-Green strain tensor corresponding to the basic 
deformation.
In the particular cases where $n=\pm 1$, we determine the 
corresponding forms of the strain-energy function. 
In every case, the slownesses are presented explicitly.
Finally we present the very simple results when 
$n = \textstyle{\frac{1}{2}}$.

\section{Inhomogeneous longitudinal plane waves.}

Here we recall some basic properties of inhomogeneous plane waves and 
then introduce inhomogeneous \textit{longitudinal} plane waves. 
Finally we introduce inhomogeneous longitudinal \textit{circularly} 
polarized plane waves.

The displacement field corresponding to an infinite train of 
infinitesimal inhomogeneous plane waves is $\epsilon \mathbf{u}$ 
(where $\epsilon$ is an infinitesimal constant: 
$\epsilon^2 \ll |\epsilon|$) and $\mathbf{u}$ is given by 
\begin{align} 
\mathbf{u} & = [ \mathbf{A} \exp i \omega (\mathbf{S \cdot x} - t) ]^+ 
\nonumber \\ 
    & = \{ \mathbf{A^+} \cos \omega (\mathbf{S^+ \cdot x} - t) 
            - \mathbf{A^-} \sin \omega (\mathbf{S^+ \cdot x} - t) \} 
                        \exp - \omega \mathbf{S^- \cdot x}. 
\end{align}
Here $\mathbf{A} = \mathbf{A^+} + i \mathbf{A^-}$ is the amplitude 
bivector; $\mathbf{S} = \mathbf{S^+} + i \mathbf{S^-}$ is the slowness 
bivector; and the period of the waves is $2\pi / \omega$. 
The particle initially at $\mathbf{x}$ is displaced to 
$\mathbf{x} + \epsilon \mathbf{u}$ at time $t$, so that it moves on an 
ellipse, centre $\mathbf{x}$, with conjugate radii 
$\epsilon  \mathbf{A^+} \exp - \omega \mathbf{S^- \cdot x}$ and 
$\epsilon  \mathbf{A^-} \exp - \omega \mathbf{S^- \cdot x}$. 
When the particle is at 
$\mathbf{x} + \epsilon \mathbf{A^+} \exp -\omega \mathbf{S^- \cdot x}$
it is moving parallel to $\mathbf{A^-}$, and when it is at  
$\mathbf{x} + \epsilon \mathbf{A^-} \exp -\omega \mathbf{S^- \cdot x}$ 
it is moving parallel to $\mathbf{A^+}$.
The sense of description of the ellipse is from the tip of 
$\mathbf{x} +\epsilon \mathbf{A^+} \exp - \omega \mathbf{S^- \cdot x}$
towards the tip of 
$\mathbf{x} +\epsilon \mathbf{A^-} \exp -\omega \mathbf{S^- \cdot x}$ 
that is, from the tip of $\mathbf{A^+}$ towards the tip of 
$\mathbf{A^-}$. 

It has been pointed out \cite{Haye84} that the slowness $\mathbf{S}$ 
may not be prescribed a priori. 
Rather, $\mathbf{S}$ is written as 
\begin{equation}
\mathbf{S} = N \mathbf{C} 
 = T e^{i \phi} (m \mathbf{\hat{m}} + i \mathbf{\hat{n}}), 
\end{equation} 
where 
\begin{equation}
\mathbf{C}  =  m \mathbf{\hat{m}} + i \mathbf{\hat{n}}, 
\quad m \ge 1, 
\quad 
\mathbf{\hat{m} \cdot \hat{n}} =0, 
\quad 
|\mathbf{\hat{m}}| = |\mathbf{\hat{n}}| = 1.
\end{equation} 
In the directional ellipse approach \cite{Haye84}, what is prescribed 
is the directional bivector $\mathbf{C}$, or equivalently, the 
directional ellipse associated with $\mathbf{C}$ that is, 
$(\mathbf{\hat{m} \cdot x})^2 / m^2 +(\mathbf{\hat{n} \cdot x})^2 =1$, 
$\mathbf{\hat{m}} \times \mathbf{\hat{n}} \mathbf{ \cdot x} = 0$.  
On choosing $\mathbf{C}$, then $N$, and thus 
$\mathbf{S} = N \mathbf{C}$, are determined from the equations of 
motion, and $\mathbf{A}$ follows as an eigenbivector of the acoustical 
tensor \cite{Haye84}.
Let $\mathbf{\hat{p}} = \mathbf{\hat{m}} \times \mathbf{\hat{n}}$ be 
the unit normal to the plane of the directional bivector $\mathbf{C}$. 
There is an infinity of choices of directional ellipses, first by 
varying the magnitude of $m$, keeping 
$\mathbf{\hat{m}}$ and $\mathbf{\hat{n}}$ held fixed and orthogonal, 
and then by rotating ($\mathbf{\hat{m}}, \mathbf{\hat{n}}$) into 
another orthogonal pair  ($\mathbf{\hat{m}}^*, \mathbf{\hat{n}}^*$) 
(say) with 
$\mathbf{\hat{p}} = \mathbf{\hat{m}}^* \times \mathbf{\hat{n}}^*$
and repeating the procedure. 
Finally $\mathbf{\hat{p}}$ is rotated and the process continued. 
In this way every possible inhomogeneous plane wave solution is 
obtained. 

For a `longitudinal' inhomogeneous plane wave the displacement field 
$\epsilon \mathbf{u}$ may be written
\begin{equation}
\mathbf{u} = [\delta \mathbf{S}
  \exp i \omega (\mathbf{S \cdot x} -t)]^+,
\end{equation}
where $\delta$ is a constant. 
Writing 
\begin{equation} 
\delta = |\delta| e^{i\psi}, \quad 
\mathbf{S} = N\mathbf{C}  =  N(m \mathbf{\hat{m}} + i \mathbf{\hat{n}})
=Te^{i\phi}(m \mathbf{\hat{m}} + i \mathbf{\hat{n}}), 
\end{equation} 
we find 
\begin{multline} \label{ellipse}
\mathbf{u} = 
   |\delta| T \exp - \omega T [(\sin \phi) m \mathbf{\hat{m}\cdot x} 
                      + (\cos \phi)  \mathbf{\hat{n} \cdot x}] \\ 
\times
 \{m \mathbf{\hat{m}} 
       \cos [ \omega T(\cos \phi) m \mathbf{\hat{m}\cdot x} 
                - \omega T(\sin \phi)  \mathbf{\hat{n} \cdot x} 
                                 - \omega t + \psi + \phi ] \\
- \mathbf{\hat{n}} 
        \sin [ \omega T(\cos \phi) m \mathbf{\hat{m}\cdot x} 
              -  \omega T(\sin \phi)  \mathbf{\hat{n} \cdot x} 
                              - \omega t + \psi + \phi] \}.
\end{multline}
The planes of constant phase are
\begin{equation}
(\cos \phi) m \mathbf{\hat{m}\cdot x} 
           - (\sin \phi)  \mathbf{\hat{n} \cdot x} =\text{constant},
\end{equation}
and the planes of constant amplitude are
\begin{equation}
(\sin \phi) m \mathbf{\hat{m}\cdot x} 
           + (\cos \phi) \mathbf{\hat{n} \cdot x} = \text{constant}.
\end{equation}
The particle paths are ellipses with principal axes along 
$\mathbf{\hat{m}}$ and $\mathbf{\hat{n}}$. 
We note that 
\begin{align}
& \mathbf{u}(\mathbf{0},t) = 
 |\delta| T [ m \mathbf{\hat{m}} \cos (\psi + \phi - \omega t) 
                   - \mathbf{\hat{n}} \sin (\psi + \phi - \omega t)], 
\nonumber \\
& \mathbf{u}(\mathbf{0}, \omega t = \psi + \phi ) = 
 |\delta| T  m \mathbf{\hat{m}}, 
\nonumber \\
& \mathbf{u}(\mathbf{0}, \omega t = \psi + \phi  + \pi /2) = 
 |\delta| T   \mathbf{\hat{n}}.
\end{align} 
Hence the sense of description for the motion of the particle 
initially at $\mathbf{x} = \mathbf{0}$ and now on the ellipse 
\eqref{ellipse} with center at  $\mathbf{x} = \mathbf{0}$, is from the 
tip of the principal axis along $\mathbf{\hat{m}}$ towards the tip of 
the principal axis along $\mathbf{\hat{n}}$.
This is clearly the case for the motions of all other particles also.

For circularly polarized waves we take $m=1$ so that the planes of 
constant phase are orthogonal to the planes of constant amplitude. 
Also, 
\begin{multline} \label{circle}
\mathbf{u} = 
   |\delta| T \exp - \{\omega T [(\sin \phi)  \mathbf{\hat{m}} 
              + (\cos \phi)  \mathbf{\hat{n}}] \mathbf{ \cdot x}\}
\\
\times [ \mathbf{\hat{m}} \cos (\kappa -  \omega t)  
    -  \mathbf{\hat{n}} \sin (\kappa - \omega t) ],
\end{multline}
where 
\begin{equation}
\kappa =  \omega T[(\cos \phi)  \mathbf{\hat{m}} 
   - (\sin \phi)  \mathbf{\hat{n}}] \mathbf{ \cdot x} + \psi + \phi. 
\end{equation}
The radius of the circle of polarization for the particle initially at 
$\mathbf{x}$ is
\begin{equation}
|\epsilon  \delta| T \exp - \{\omega T [(\sin \phi)  \mathbf{\hat{m}} 
                + (\cos \phi)  \mathbf{\hat{n}}] \mathbf{ \cdot x}\}.
\end{equation}
The displacement $\epsilon \mathbf{u}$ where $\mathbf{u}$ is given by 
\eqref{circle} corresponds to an infinite train of circularly 
polarized longitudinal inhomogeneous plane waves.

\section{Basic equations.}

\subsection{Hyperelastic materials.}
\label{Hyperelastic materials}

We consider a body made of homogeneous isotropic elastic 
material, initially in a state of rest $\mathcal{B}_0$.
When the body is deformed to a state  $\mathcal{B}$ (say), a
material particle which was at $\mathbf{X}$ in $ \mathcal{B}_0$ moves
to the position $\mathbf{x}$, at time $t$, where
  \begin{equation} \label{xDef}
\mathbf{x} = \mathbf{x} ( \mathbf{X},t).
\end{equation}
The left Cauchy-Green strain tensor $\mathbb{B}$ is defined by
 \begin{equation}\label{Bdef}
\mathbb{B}_{ij}=
(\partial x_i / \partial X_A) (\partial x_j /\partial X_A),
\end{equation} 
in a rectangular Cartesian coordinate system, fixed in $ \mathcal{B}$.

Three principal invariants of $\mathbb{B}$ are 
$\text{I}, \text{II}, \text{III}$ defined by 
 \begin{equation} \label{invDef}
\text{I}= \text{ tr } \mathbb{B}, 
\quad 2\text{II}= \text{I}^2 - \text{ tr } (\mathbb{B}^2),
\quad \text{III}= \text{ det } \mathbb{B}.
\end{equation}  

The constitutive equation for the material, relating the Cauchy stress 
tensor $\mathbb{T}$ with the strain tensor $ \mathbb{B}$ is 
\cite{CuHa69}
 \begin{equation} \label{Tdef}
\mathbb{T} = N_0 \mathbf{1}
+ N_1 \mathbb{B} - N_{-1} \mathbb{B}^{-1},
\end{equation}  
where the material coefficients $N_\Gamma$  ($\Gamma = 0, \pm 1$) are
functions of $\text{I}, \text{II}, \text{III}$.
In the case of a hyperelastic material, the strain energy density 
function $W$, measured per unit mass, is characteristic of the 
material and a function of $\text{I}, \text{II}, \text{III}$:
\begin{equation}
  W = W(\text{I}, \text{II}, \text{III}),
\end{equation}
and then $N_0$, $N_1$, $N_{-1}$ are given by 
\cite{CuHa69}
 \begin{align}  \label{Ndef}
& N_0 = 2[\text{II} \, \text{III}^{-\halft} \frac{\partial W}{\partial \text{II}} +
\text{III}^\halft  \frac{\partial W}{\partial \text{III}}], 
\nonumber \\ 
&  N_1 = 2
     \text{III}^{-\halft} \frac{\partial W}{\partial \text{I}} , 
\quad 
N_{-1} = 2
\text{III}^\halft \frac{\partial W}{\partial \text{II}}.
\end{align}  

\subsection{Finite static pure homogeneous deformation.}
\label{Homogeneous Deformation}

We assume that the body is first subjected to a finite triaxial
static homogeneous deformation, as it is deformed from 
$\mathcal{B}_0$ to $\mathcal{B}$. 
Then, the corresponding three principal axes of
deformation constitute a fixed rectangular Cartesian coordinate
system ($O, \mathbf{i}, \mathbf{j}, \mathbf{k}$) say, in which the
deformation \eqref{xDef} is written as
 \begin{equation} \label{xHom}
\mathbf{x}= \lambda_1 X \mathbf{i} + \lambda_2 Y \mathbf{j} +
 \lambda_3 Z \mathbf{k},
\end{equation}  
where the $\lambda$'s are the principal stretch ratios, assumed,
without loss of generality, to be ordered as
 \begin{equation}\label{order}
  \lambda_1 > \lambda_2 > \lambda_3.
\end{equation}  

In this case, the left Cauchy--Green tensor reduces to
 \begin{equation} \label{B}
\mathbb{B} =  \lambda_1^2 \mathbf{i} \otimes \mathbf{i}
+ \lambda_2^2 \mathbf{j}\otimes \mathbf{j}
+ \lambda_3^2 \mathbf{k} \otimes \mathbf{k},
\end{equation}  
with corresponding invariants
 \begin{equation} \label{invHom}
\text{I} =  \lambda_1^2 + \lambda_2^2 + \lambda_3^2, 
\quad \text{II} =  \lambda_1^2
\lambda_2^2 + \lambda_2^2  \lambda_3^2 + \lambda_3^2 \lambda_1^2, 
\quad
\text{III} =  \lambda_1^2 \lambda_2^2 \lambda_3^2.
\end{equation}  
The Cauchy stress necessary to maintain the basic finite static 
homogeneous deformation \eqref{xHom} has components
 \begin{align}  \label{THom}
& T_{\alpha \beta}=  0, \quad \alpha \ne \beta \nonumber \\
& T_{\alpha \alpha}= N_0 
 + N_1 \lambda_\alpha^2 - N_{-1}\lambda_\alpha^{-2},
\end{align}  
where the $N_\Gamma$ ($\Gamma = -1,0,1$) are defined as in
\eqref{Ndef}, at values of $\text{I}, \text{II}, \text{III}$ given by 
\eqref{invHom}.

\subsection{Superposed `longitudinal' inhomogeneous plane\\ waves.}
\label{Superposed longitudinal inhomogeneous plane  waves}

Let the body be subjected to a further deformation bringing it 
from the state $\mathcal{B}$ to the state $\overline{\mathcal{B}}$,
corresponding to the propagation of an infinitesimal
\textit{`longitudinal' inhomogeneous plane wave} of complex
exponential type in the deformed material, such that the current
position $\overline{\mathbf{x}}$ in $\overline{\mathcal{B}}$, of a
particle at $\mathbf{x}$ given by \eqref{xHom} in $\mathcal{B}$, and at
$\mathbf{X}$ in $\mathcal{B}_0$, is given by
 \begin{equation} \label{xBar}
\mathbf{\overline{x}} = \mathbf{x}+ \epsilon \{ \mathbf{S} e^{i
\omega(\mathbf{S \cdot x}-t)} \}^+.
\end{equation}  
Here, $\epsilon$ is a small parameter, such that terms of order higher
than $\epsilon$ may be neglected in comparison with first order terms. 

Now, corresponding to the superposition of the small-amplitude motion
\eqref{xBar} upon the large deformation \eqref{xHom} are the quantities
$\overline{\mathbb{B}}$, $\overline{\mathbb{B}}^{-1}$, 
$\overline{\text{I}}$, $\overline{\text{II}}$,
$\overline{\text{III}}$, $\overline{N}_{-1}$, $\overline{N}_0$,
$\overline{N}_1$, and $\overline{\mathbb{T}}$, that is, respectively,
the left Cauchy-Green tensor and its
inverse, the first three invariants, the three material constitutive
parameters, and the Cauchy stress tensor. Each of these quantities may
be expanded up to the first order in $\epsilon$, and written in the
form,
 \begin{align}
& \overline{\mathbb{B}} = 
  \mathbb{B} + \epsilon \{ i \omega \mathbb{B}^*
                   e^{i \omega(\mathbf{S \cdot x}-t)} \}^+, \dots
\nonumber \\ 
& \overline{\text{I}} = \text{tr }\overline{\mathbb{B}} = 
  \text{I} + \epsilon \{ i \omega \text{I}^*
                   e^{i \omega(\mathbf{S \cdot x}-t)} \}^+,  \dots
\nonumber \\ 
& \overline{\mathbb{T}} =
  \mathbb{T} + \epsilon \{ i \omega \mathbb{T}^* 
                   e^{i \omega(\mathbf{S \cdot x}-t)} \}^+.
\end{align}  
Here, we have,
 \begin{align}  \label{incrementalQties}
&\mathbb{B}^* = \mathbf{S} \otimes \mathbb{B}\mathbf{S} +
   \mathbb{B}\mathbf{S} \otimes \mathbf{S}, \nonumber 
\\
&(\mathbb{B}^{-1})^* = -\mathbf{S} \otimes \mathbb{B}^{-1}\mathbf{S} 
  - \mathbb{B}^{-1}\mathbf{S} \otimes \mathbf{S}, \nonumber 
\\ 
& \text{I}^* = 2(\mathbf{S \cdot} \mathbb{B} \mathbf{S}),  \nonumber 
\\ 
& \text{II}^* = 2[\text{II}(\mathbf{S \cdot S}) 
  - \text{III}(\mathbf{S \cdot} \mathbb{B}^{-1}\mathbf{S})], 
\\ 
& \text{III}^* = 2 \text{III} (\mathbf{S \cdot S}),  \nonumber 
\\ 
& N^*_\Gamma = \text{I}^* N_{\Gamma, \text{I}} 
  + \text{II}^* N_{\Gamma, \text{II}} 
     + \text{III}^*N_{\Gamma, \text{III}}, 
    \quad (\Gamma = -1,0,1), \nonumber  
\\  
& \mathbb{T}^* = N^*_0 \mathbf{1} + N_1^* \mathbb{B} 
  - N_{-1}^* \mathbb{B}^{-1} +  N_1 \mathbb{B}^* 
   - N_{-1} (\mathbb{B}^{-1})^*.  \nonumber
\end{align}  

Finally, the mass density  $\overline{\rho}$ in the current
configuration is
 \begin{equation} \label{rhoBar}
\overline{\rho} = \text{III}^{-\halft} 
      \rho ( 1 - \epsilon \{ i \omega (\mathbf{S \cdot S}) 
                       e^{i \omega ( \mathbf{S \cdot x}-t)}\}^+).
\end{equation}

\subsection{Equations of motion.}
\label{Equations of motion}

In the dynamic state $\overline{\mathcal{B}}$, the equations of motion,
written in the absence of body forces, read
 \begin{equation}\label{eqMotion}
\frac{\partial \overline{T}_{ij}}{\partial \overline{x}_j} =
\overline{\rho} \frac{\partial^2 \overline{x}_i}{\partial t^2}.
\end{equation}  
Up to the first order in $\epsilon$, they are equivalent to
$\mathbb{T}^* \mathbf{\cdot S} = \rho \mathbf{S}$, or
 \begin{equation} \label{S,BS,B(-1)S}
\Theta \mathbf{S}
  + \Phi \mathbb{B} \mathbf{S} 
   +  \Gamma  \mathbb{B}^{-1} \mathbf{S} 
  = \mathbf{0},
\end{equation}  
where 
\begin{align} 
& \Theta := N^*_0 + N_1(\mathbf{S \cdot} \mathbb{B} \mathbf{S}) 
 + N_{-1}(\mathbf{S \cdot} \mathbb{B}^{-1} \mathbf{S}) - \rho,
\nonumber  \\ 
& \Phi := N^*_1 + N_1(\mathbf{S \cdot S}), 
\nonumber \\ 
& \Gamma := N^*_{-1} - N_{-1}(\mathbf{S \cdot S}).
\end{align}  
When referred to axes $\mathbf{i}$, $\mathbf{j}$, $\mathbf{k}$, 
the principal axes of $\mathbb{B}$, Eqs.~\eqref{S,BS,B(-1)S} may 
be written
 \begin{equation} 
(\Theta + \Phi \lambda_\alpha^2 + \Gamma \lambda_\alpha^{-2})C_\alpha
  = 0, \quad \alpha = 1,2,3, \text{ no sum},
\end{equation}  
where $\mathbf{C} = C_1 \mathbf{i} + C_2 \mathbf{j} + C_3\mathbf{k}$.
These equations have to be satisfied for all choices of $C_1$, $C_2$, 
$C_3$, and for all positive choices of $\lambda_1$, $\lambda_2$, 
$\lambda_3$ satisfying \eqref{order}. 
Noting that 
\begin{equation}
  \begin{vmatrix}
    1 & \lambda_1^2 & \lambda_1^{-2} \\
    1 & \lambda_2^2 & \lambda_2^{-2} \\
    1 & \lambda_3^2 & \lambda_3^{-2} 
  \end{vmatrix} = 
  (\lambda_1^2 - \lambda_2^2)
    (\lambda_2^2 - \lambda_3^2)
      (\lambda_3^2 - \lambda_1^2)/\text{III} \ne 0,
\end{equation}
it follows that we must have
\begin{equation}
 \Theta =  \Phi =  \Gamma = 0. 
\end{equation}
Thus, writing $\mathbf{S} = N \mathbf{C}$, we have
\begin{multline} \label{3rdEqn}
\rho N^{-2} = (2 N_{0,\text{I}} + N_1) 
         \mathbf{C \cdot} \mathbb{B} \mathbf{C} 
 + (-2\text{III} N_{0,\text{II}} 
       + N_{-1}) \mathbf{C \cdot} \mathbb{B}^{-1} \mathbf{C} \\
 + (2\text{II} N_{0,\text{II}} + 2 \text{III} N_{0,\text{III}}) 
                                                  \mathbf{C \cdot C},
\end{multline}   
and also
 \begin{align}  \label{eqMotion2}
& 2N_{1,\text{I}}(\mathbf{C \cdot} \mathbb{B} \mathbf{C}) 
 - 2\text{III} N_{1,\text{II}} 
        (\mathbf{C \cdot} \mathbb{B}^{-1} \mathbf{C})
  \nonumber
\\ 
& \phantom{12345} 
 + [N_1 + 2 \text{II} N_{1,\text{II}} 
     + 2 \text{III} N_{1,\text{III}}](\mathbf{C \cdot C}) = 0, 
\\
& 2N_{-1,\text{I}}(\mathbf{C \cdot} \mathbb{B} \mathbf{C}) 
   - 2\text{III} N_{-1,\text{II}} 
        (\mathbf{C \cdot} \mathbb{B}^{-1} \mathbf{C}) 
\nonumber 
\\
& \phantom {12345}
 +  [-N_{-1} + 2 \text{II} N_{-1,\text{II}}
  + 2 \text{III} N_{-1,\text{III}}](\mathbf{C \cdot C}) = 0. 
\label{eqMotion3}
\end{align}  
Here $\mathbf{C}$ is assumed prescribed. 
Then, $\mathbf{S} = N \mathbf{C}$ is determined from \eqref{3rdEqn}.
The response functions $N_1$, $N_{-1}$ must be such that 
\eqref{eqMotion2} and \eqref{eqMotion3} are satisfied for all choices 
of $\mathbf{C}$ and all positive choices of  $\lambda_1$, 
$\lambda_2$, $\lambda_3$.

\section{Propagation of `longitudinal' in\-ho\-mo\-ge\-ne\-ous 
plane waves  for any directional
bivector $\mathbf{C}$.}  
\label{Propagation of longitudinal inhomogeneous plane waves}

Here we find the most general form of the stored energy density for 
which infinitesimal longitudinal inhomogeneous plane waves 
may propagate in the finitely deformed material for any choice of 
directional bivector $\mathbf{C}$. 

\subsection{`Longitudinal' waves of general  polarization.}
\label{Longitudinal waves of general  polarization}

Equations \eqref{eqMotion2} and \eqref{eqMotion3} must be satisfied 
for any $\mathbf{C}$. 
In particular, choose $\mathbf{C} \ne \mathbf{0}$ such that 
$\mathbf{C \cdot C} = 0$,  
$\mathbf{C \cdot} \mathbb{B}^{-1} \mathbf{C} = 0$. 
In that case, $\mathbf{C \cdot } \mathbb{B} \mathbf{C} \ne 0$
and it follows from \eqref{eqMotion2}, \eqref{eqMotion3} that
 \begin{equation} 
N_{1,\text{I}} = N_{-1,\text{I}} = 0. 
\end{equation}  
Next choose $\mathbf{C} \ne \mathbf{0}$ such that 
$\mathbf{C \cdot C} = 0$,  
$\mathbf{C \cdot} \mathbb{B} \mathbf{C} = 0$. 
In that case, $\mathbf{C \cdot } \mathbb{B}^{-1} \mathbf{C} \ne 0$
and it follows that
 \begin{equation} 
N_{1,\text{II}} = N_{-1,\text{II}} = 0. 
\end{equation}  
Thus
 \begin{equation} \label{N(III)}
 N_{\Gamma} = N_{\Gamma}(\text{III}), \quad \Gamma
= -1,1.
\end{equation}  

The $N_{\Gamma}$ are functions of $\text{III}$ alone, and equations
\eqref{eqMotion2} and \eqref{eqMotion3} reduce to
 \begin{equation}
 [N_1 +  2 \text{III} N_{1}'](\mathbf{C \cdot C}) = 0, \quad
 [-N_{-1} +  2 \text{III} N_{-1}'](\mathbf{C \cdot C}) =0.
\end{equation}  
These equations must be satisfied with any $\mathbf{C}$, in particular
with $\mathbf{C}$ such that $\mathbf{C \cdot C} \ne 0$ 
(e.g. $\mathbf{C} = \mathbf{i} + 2i \mathbf{j}$). So,
 \begin{equation}
 N_1 +  2 \text{III} N_{1}'= 0, \quad
 -N_{-1} +  2 \text{III} N_{-1}' =0,
\end{equation}  
and by integration,
 \begin{equation} \label{NforHadamard}
 N_1  = 2 \nu \text{III}^{-\halft}, \quad
 N_{-1} = 2 \mu \text{III}^{\halft},
\end{equation}  
where $\nu$ and $\mu$ are two constant parameters, independent of the
strain invariants. The corresponding hyperelastic material has the
following constitutive equation, by \eqref{Tdef},
 \begin{equation}\label{HadamardConstitutive}
\mathbb{T} = N_0(\text{I},\text{II},\text{III}) \mathbf{1} 
+ 2\nu \text{III}^{-\halft} \mathbb{B} 
 - 2 \mu \text{III}^{\halft} \mathbb{B}^{-1}.
\end{equation}  
In order that this expression should correspond to that of an 
hyperelastic material, it is seen that $N_0$ has to be independent of 
$\text{I}$ and linear in $\text{II}$, thus
\begin{equation} \label{N0hyperelastic}
N_0 = 2 \left( \text{II } \text{III}^{-\halft} \mu 
      +  \text{III}^{\halft} \dfrac{\text{d}h}{\text{d III}} \right),
\end{equation}
so that the strain energy density is given by
\begin{equation} \label{WforHadamard}
W = \nu \text{I} + \mu \text{II} + h(\text{III}),
\end{equation}  
characteristic of the Hadamard material \cite{John66}.
Here $h$ is an arbitrary function of $\text{III}$.  
We conclude:
\begin{quote}
The most general material in which `longitudinal' 
\underline{inhomogeneous} plane waves may propagate for any choice of 
the directional bivector $\mathbf{C}$, when it is held in any state of 
finite static pure homogeneous deformation, is the Hadamard material.
\end{quote}
This same conclusion was reached by John \cite{John66} for the 
propagation of finite amplitude \textit{homogeneous} plane waves.

Using \eqref{3rdEqn} and \eqref{NforHadamard}, we note that the 
slowness $\mathbf{S}$ corresponding to the directional bivector 
$\mathbf{C}$ is given by $\mathbf{S} = N \mathbf{C}$, where 
\begin{multline} \label{Nhadamard}
\rho N^{-2} = 
  2(N_{0,\text{I}} 
   + \nu \text{III}^{-\halft})\mathbf{C \cdot} \mathbb{B}\mathbf{C} 
+ 2( - \text{III}N_{0,\text{II}} + \mu \text{III}^{\halft}) 
     \mathbf{C \cdot} \mathbb{B}^{-1}\mathbf{C}\\
+ 2(\text{II}N_{0,\text{II}} 
       + \text{III} N_{0,\text{III}})\mathbf{C \cdot C}.
\end{multline}

In the particular case of a hyperelastic Hadamard material, on using 
\eqref{N0hyperelastic} and \eqref{WforHadamard}, equation 
\eqref{Nhadamard} becomes 
\begin{multline} \label{NhadamardHyperelastic}
\rho N^{-2} = 
  2 \nu \text{III}^{-\halft} \mathbf{C \cdot} \mathbb{B}\mathbf{C} 
   - 2 \mu \text{III}^{\halft} 
         \mathbf{C \cdot} \mathbb{B}^{-1}\mathbf{C}\\
    + 4[ \mu \text{II } \text{III}^{-\halft}
     +  \text{III} (\text{III}^\halft h')' ]\mathbf{C \cdot C}.
\end{multline}

For any choice of the directional bivector $\mathbf{C}$, whether it is 
taken to be linear, elliptic, or circular, the corresponding wave 
train is linearly, elliptically, or circularly polarized, 
respectively, the displacement field being given by 
\begin{equation} \label{displField}
\mathbf{\overline{x}} = \mathbf{x}
 + \epsilon \{ N \mathbf{C} e^{i \omega(N\mathbf{C \cdot x}-t)} \}^+.
\end{equation}  
wherein $N$ is given by \eqref{Nhadamard}.

\subsubsection*{Example}
As an example we may choose $\mathbf{C}$ such that 
\begin{equation} \label{example}
  \dfrac{C_1^2}{C_3^2} = 
 \left(\dfrac{\lambda_2^4 - \lambda_3^4}{\lambda_1^4 - \lambda_2^4}
    \right) \left(\dfrac{\lambda_1^2}{\lambda_3^2}\right), 
\quad 
 \dfrac{C_2^2}{C_3^2} = 
 -\left(\dfrac{\lambda_1^4 - \lambda_3^4}{\lambda_1^4 - \lambda_2^4}
    \right) \left(\dfrac{\lambda_2^2}{\lambda_3^2}\right). 
\end{equation}
Then 
\begin{equation} \label{condExample}
  \mathbf{C \cdot} \mathbb{B} \mathbf{C} = 
      \mathbf{C \cdot } \mathbb{B}^{-1} \mathbf{C} = 0,
\end{equation}
and  \eqref{NhadamardHyperelastic} gives
\begin{equation}
\rho N^{-2} = 
  - 4[ \mu \text{II } \text{III}^{-\halft}
     +  \text{III} (\text{III}^\halft h')' ]
   \dfrac{(\lambda_1^2 - \lambda_3^2) 
     (\lambda_2^2 - \lambda_3^2) 
       (\lambda_2^2 - \lambda_3^2)}
         { \lambda_3^2 (\lambda_1^2 + \lambda_2^2)} C_3^2 .
\end{equation}

Thus $N C_3$ is determined in terms of $\lambda_\alpha$, $h'$, 
and $\mu$, and from \eqref{example}, $NC_1$ and $NC_2$ are also 
determined. 
Thus the displacement field \eqref{displField} may be written 
explicitly in terms of the basic deformation stretches 
$\lambda_\alpha$, and $\mu$, $h'$.

We note in passing that \eqref{condExample} may be interpreted 
\cite{BoHa93} in terms of the two special central planes for which 
the sections of the ellipsoids 
$\mathbf{x \cdot } \mathbb{B} \mathbf{x} = 1$, 
$\mathbf{x \cdot } \mathbb{B}^{-1} \mathbf{x} = 1$, 
are a pair of similar and similarly situated ellipses. 
The plane of $\mathbf{C}$ must coincide with either of the two 
special central planes and also the ellipse of $\mathbf{C}$ is 
similar and similarly situated to the elliptical sections of the 
$\mathbb{B}$ and $\mathbb{B}^{-1}$ ellipsoids by the plane of 
$\mathbf{C}$.

\subsection{`Longitudinal' waves of circular polarization.}
\label{Longitudinal waves of circular polarization}

Now we determine the most general material for which infinitesimal 
`longitudinal' inhomogeneous plane waves of \textit{circular} 
polarization may propagate for any choice of isotropic directional 
bivector $\mathbf{C}$, when the material is held in an arbitrary state 
of finite static pure homogeneous deformation.
Of course, as we have seen already, such waves may propagate in a 
deformed Hadamard material.

The condition that such waves may propagate is that equations 
\eqref{eqMotion2} and \eqref{eqMotion3} be satisfied for all
isotropic $\mathbf{C}$: $\mathbf{C \cdot C} = 0$ and for all positive 
choice of $\lambda_1$, $\lambda_2$, $\lambda_3$.
When  $\mathbf{C \cdot C} = 0$,  equations 
\eqref{eqMotion2} and \eqref{eqMotion3}  reduce to
 \begin{equation}
N_{\Gamma,\text{I}}(\mathbf{C \cdot} \mathbb{B} \mathbf{C})
 - \text{III} N_{\Gamma,\text{II}} 
        (\mathbf{C \cdot} \mathbb{B}^{-1} \mathbf{C}) = 0, \quad
(\Gamma = -1,1).
\end{equation}  
Because $\lambda_1$, $\lambda_2$, $\lambda_3$ are arbitrary, it 
follows that 
\begin{equation}
N_{\Gamma, \text{I}} = N_{\Gamma, \text{II}} = 0,\quad
(\Gamma = -1,1),
\end{equation}
so that the corresponding constitutive equation is
 \begin{equation} \label{constitutiveGeneralized}
\mathbb{T} = N_0(\text{I},\text{II},\text{III}) \mathbf{1} 
 + N_1(\text{III}) \mathbb{B} - N_{-1}(\text{III}) \mathbb{B}^{-1}.
\end{equation}  
The corresponding materials are called 
'Generalized Hadamard materials'.
The class encompasses Hadamard materials.

For any choice of isotropic $\mathbf{C}$, the wave train is circularly 
polarized, with the plane of $\mathbf{C}$ being the plane of 
polarization, and the corresponding slowness 
$\mathbf{S} = N \mathbf{C}$ being given through  
\begin{equation} \label{NGeneralHadamard}
\rho N^{-2} = 
  (2N_{0,\text{I}} + N_1)\mathbf{C \cdot} \mathbb{B}\mathbf{C} 
+ (N_{-1} - 2\text{III}N_{0,\text{II}}) 
     \mathbf{C \cdot} \mathbb{B}^{-1}\mathbf{C},
\end{equation}  
on using \eqref{3rdEqn}.

If the Generalized Hadamard material is to be hyperelastic, then it 
may be shown that $N_0$ must have the form 
\begin{equation}
N_0 = \text{I } \text{III}^\halft (\text{III}^{\halft}N_1)' 
  - \text{II } \text{III}^{-\halft} (\text{III}^{\halft}N_{-1})' 
   + \text{III}^{\halft} h'(\text{III}),
\end{equation}
and the corresponding form of the strain energy density $W$ is 
\begin{equation}
2W =
 \text{I } \text{III}^{\halft}N_1(\text{III})
  + \text{II }\text{III}^{-\halft}N_{-1}(\text{III}) 
   + h(\text{III}),
\end{equation}  
where $N_1$, $N_{-1}$, and $h$ are arbitrary functions of $\text{III}$.

\subsubsection*{Remark: A universal relation.}
Using equation \eqref{NGeneralHadamard} we derive a universal relation
among the wave slownesses corresponding to three choices of the 
directional bivector $\mathbf{C}$.
We write $\mathbf{C_1} = \mathbf{i} + i \mathbf{j}$, 
$\mathbf{C_2} = \mathbf{j} + i \mathbf{k}$, 
$\mathbf{C_3} = \mathbf{k} + i \mathbf{j}$. 
Then $\mathbf{C_\alpha \cdot C_\alpha} = 0$, and 
\begin{align}
& \mathbf{C_1 \cdot} \mathbb{B}\mathbf{C_1}
  = \lambda_1^2 - \lambda_2^2, 
&& \mathbf{C_1 \cdot} \mathbb{B}^{-1}\mathbf{C_1}
  = \lambda_1^{-2} - \lambda_2^{-2}, 
\nonumber \\ 
&  \mathbf{C_2 \cdot} \mathbb{B}\mathbf{C_2}
  = \lambda_2^2 - \lambda_3^2, 
&&
\mathbf{C_2 \cdot} \mathbb{B}^{-1}\mathbf{C_2}
  = \lambda_2^{-2} - \lambda_3^{-2}, 
\nonumber \\ 
& \mathbf{C_3 \cdot} \mathbb{B}\mathbf{C_3}
  = \lambda_3^2 - \lambda_1^2, 
&&
\mathbf{C_3 \cdot} \mathbb{B}^{-1}\mathbf{C_3}
  = \lambda_3^{-2} - \lambda_2^{-2}. 
\end{align}
Let the value of $N$ corresponding to $\mathbf{C_\alpha}$ be 
$N(\mathbf{C_\alpha})$. 
Then 
\begin{equation} 
\rho N^{-2}(\mathbf{C_1}) = 
  (2N_{0,\text{I}} + N_1)(\lambda_1^2 - \lambda_2^2) 
+ (N_{-1} - 2\text{III}N_{0,\text{II}}) 
    (\lambda_1^{-2} - \lambda_2^{-2}),
\end{equation}  
etc., and we have the universal relation
\begin{equation}
N^{-2}(\mathbf{C_1}) + N^{-2}(\mathbf{C_2}) + N^{-2}(\mathbf{C_3}) = 0.
\end{equation}

\section{Propagation of `longitudinal' inhomogeneous \\
plane waves, circularly polarized in special planes} 
\label{special planes}

Now we seek elastic isotropic materials which allow the propagation of
two very special circularly polarized `longitudinal' inhomogeneous 
plane waves, irrespective of the basic static homogeneous deformation. 
The circle of polarization of the waves is to be one or the other of 
the two central sections of the ellipsoid 
$ \mathbf{x \cdot} \mathbb{B}^n \mathbf{x} = 1$ 
(the $\mathbb{B}^n$-ellipsoid), where 
$\mathbb{B}$ is the left Cauchy-Green strain tensor \eqref{B} 
corresponding to the basic static deformation.

We first consider the cases where $n=1$ and where $n=-1$, 
and then treat the general case.

\subsection{Circle of polarization lies in a plane of central\\
circular section of the $\mathbb{B}$-ellipsoid or of
the \\$\mathbb{B}^{-1}$-ellipsoid.}
\label{central circular section of the B-ellipsoid}

The $\mathbb{B}$-ellipsoid is the surface described by 
$\mathbf{x \cdot} \mathbb{B} \mathbf{x} = 1$. 
Because $\mathbb{B}$ is a positive definite tensor, there exist two 
central planes which cut the ellipsoid in circular sections. 
Boulanger and Hayes \cite{BoHa93} have proved that
the circles are described by the bivector $ \mathbf{A}$ (say) such that
$ \mathbf{A \cdot A} =  \mathbf{A \cdot} \mathbb{B} \mathbf{A} = 0$.
Hence, the wave \eqref{xBar}, with a bivector $\mathbf{C}$ satisfying
 \begin{equation} \label{CBC=0}
 \mathbf{C \cdot C} =  \mathbf{C \cdot} \mathbb{B} \mathbf{C} = 0,
 \end{equation}  
is a `longitudinal' inhomogeneous plane wave, circularly polarized,
with a central circular section of the $\mathbb{B}$-ellipsoid as circle
of polarization. 
When \eqref{CBC=0} is satisfied, it follows that 
$\mathbf{C \cdot} \mathbb{B}^{-1} \mathbf{C} \ne 0$ unless 
$\mathbf{C} \equiv \mathbf{0}$, so that \eqref{eqMotion2}, 
\eqref{eqMotion3} reduce to
 \begin{equation}
N_{1,\text{II}} = N_{-1,\text{II}} = 0.
\end{equation}  
These must hold for all possible positive 
$\lambda_1$, $\lambda_2$, $\lambda_3$.
Hence $N_1 = N_1(\text{I}, \text{III})$, 
$N_{-1} = N_{-1}(\text{I}, \text{III})$ and the corresponding 
constitutive equation is
 \begin{equation}\label{materialCBC=0}
\mathbb{T} = N_0(\text{I},\text{II},\text{III}) \mathbf{1}
  + N_1(\text{I},\text{III}) \mathbb{B} 
   - N_{-1}(\text{I},\text{III}) \mathbb{B}^{-1}.
\end{equation}  
This is, of course, inclusive of the Generalized Hadamard materials 
class \eqref{constitutiveGeneralized}.
When \eqref{CBC=0} holds, then, from \eqref{3rdEqn}, 
\begin{equation} 
\rho N^{-2} = (- 2 \text{III} N_{0,\text{II}} + N_{-1})
           \mathbf{C \cdot} \mathbb{B}^{-1} \mathbf{C}.
\end{equation}
To determine the possible $\mathbf{C}$, we recall that the Hamiltonian
decomposition of the $\mathbb{B}$ tensor is
 \begin{equation} \label{hamilton}
\mathbb{B} = \lambda_2^2 \mathbf{1}
 + \halft (\lambda_1^2 - \lambda_3^2)
   [\mathbf{h^+}\otimes\mathbf{h^-}+\mathbf{h^-}\otimes\mathbf{h^+}],
\end{equation}  
where
 \begin{align} 
&\mathbf{h^\pm} = \delta \mathbf{i} \pm \phi \mathbf{k},
\quad
\delta=\sqrt{\frac{\lambda_1^2-\lambda_2^2}{\lambda_1^2-\lambda_3^2}},
\quad
\phi=\sqrt{\frac{\lambda_2^2-\lambda_3^2}{\lambda_1^2-\lambda_3^2}},
\quad
\delta^2+\phi^2=1,\\
&\delta^2 \lambda_3^2 + \phi^2 \lambda_1^2 = \lambda_2^2,
\quad
\delta^2 \lambda_3^{-2} + \phi^2 \lambda_1^{-2}
 -  \lambda_2^{-2} 
    = \delta^2\phi^2 (\lambda_1^2 - \lambda_3^2)^2/\text{III}.
\end{align}  
By \eqref{CBC=0} it follows, using \eqref{hamilton}, that 
$(\mathbf{C \cdot h^+})(\mathbf{C \cdot h^-})=0$.
Hence either $\mathbf{C \cdot h^+}=0$ or $\mathbf{C \cdot h^-}=0$, 
so that the only possible $\mathbf{C}$ are 
 \begin{equation} \label{CwhenCBC=0}
\mathbf{C} = \phi \mathbf{i} \pm i \mathbf{j} - \delta \mathbf{k},
\quad \text{or} \quad
\mathbf{C} = \phi \mathbf{i} \pm i \mathbf{j} + \delta \mathbf{k}.
\end{equation}  
For the first pair of possible $\mathbf{C}$, 
$\mathbf{C} = \phi \mathbf{i} + i \mathbf{j} - \delta \mathbf{k}$
and 
$\mathbf{C} = \phi \mathbf{i} - i \mathbf{j} - \delta \mathbf{k}$,
the polarization circles lie in the same plane (with normal 
$\mathbf{h^+}$) of central circular section of the 
$\mathbb{B}$-ellipsoid. 
However the two circles are described in opposite senses. 
A similar comment applies to the second pair of possible  
$\mathbf{C}$, 
$\mathbf{C} = \phi \mathbf{i} \pm i \mathbf{j} + \delta \mathbf{k}$.
The reason why there are four possibilities for $\mathbf{C}$ is that 
if $\mathbf{C}$ is a solution of \eqref{CBC=0} then so also is its 
complex conjugate $\overline{\mathbf{C}}$, because $\mathbb{B}$ is 
real and there are just two central circular sections of the 
$\mathbb{B}$-ellipsoid. 

Now, for all these four possible choices of $\mathbf{C}$, 
$\mathbf{C \cdot}\mathbb{B}^{-1} \mathbf{C}= 
  \delta^2 \lambda_3^{-2} + \phi^2 \lambda_1^{-2} - \lambda_2^{-2}
 = \delta^2\phi^2 (\lambda_1^2 - \lambda_3^2)^2/\text{III}$ and so $N$ 
is given by
 \begin{equation} \label{NwhenCBC=0}
\rho N^{-2} = 
  \delta^2\phi^2(\lambda_1^2-\lambda_3^2)^2 \text{III}^{-1}
            (N_{-1} - 2 \text{III} N_{0,\text{II}}).
\end{equation}  
We note that $\mathbf{C}$ and $N$ are determined completely by the 
basic deformation.
The planes of constant phase and the planes of constant amplitude are
orthogonal.
From \eqref{NwhenCBC=0}, $N$ is either purely real or purely
imaginary.
When $N$ is real (imaginary), the planes of constant phase (amplitude)
are $\phi x \mp \delta z=$ const. 
and the planes of constant amplitude (phase) are $y=$ const.
Of course, any scalar multiple of $\mathbf{S}$ is a possible amplitude
bivector.
The displacements corresponding to \eqref{CwhenCBC=0}$_1$ are
 \begin{equation}
\mathbf{\overline{x}} = \mathbf{x}
 + \epsilon e^{-\omega Ny} 
 \{(\phi \mathbf{i} \pm i \mathbf{j} - \delta \mathbf{k}) 
   e^{i \omega[N(\phi x - \delta z)-t]} \}^+.
\end{equation}  
Essentially, there are two possible circularly polarized longitudinal 
inhomogeneous plane waves for which 
$\mathbf{C \cdot}\mathbb{B} \mathbf{C}=0$.

Also, we note that for \eqref{materialCBC=0} to represent a 
hyperelastic material, the response function 
$N_{-1}(\text{I}, \text{III})$ must be independent of $\text{I}$. 
The response functions are then of the form
\begin{align} 
& N_{-1} = G(\text{III}), \quad 
N_0 =  \text{II} G'(\text{III}) 
   + [\text{II}/(2\text{III})]G(\text{III})
       + K(\text{I}, \text{III}), 
\nonumber \\
& N_1 =  \text{III}^{-\halft} L(\text{I}) +
  \text{III}^{-\halft} \textstyle{\int} 
           \partial K(\text{I}, \text{III})/ \partial \text{I } 
                                  \text{III}^{-\halft} \text{dIII}, 
\end{align}
with corresponding strain energy density $W$ given by 
\begin{equation} 
2W =  \textstyle{\int} L(\text{I}) \text{dI} 
  + \textstyle{\int} 
       K(\text{I}, \text{III}) \text{III}^{-\halft} \text{dIII}
   +  \text{II } \text{III}^{-\halft} G(\text{III}), 
\end{equation}
where $G$, $K$, $L$ are arbitrary functions of their arguments.

Similarly, the wave \eqref{xBar}, with a bivector $\mathbf{C}$ 
satisfying
 \begin{equation} \label{CB(-1)C=0}
 \mathbf{C \cdot C} =  \mathbf{C \cdot} \mathbb{B}^{-1} \mathbf{C} = 0,
 \end{equation}  
is a `longitudinal' inhomogeneous plane wave,
circularly polarized, with
a central circular section of the $\mathbb{B}^{-1}$-ellipsoid as
circle of polarization.
When \eqref{CB(-1)C=0} is satisfied, then \eqref{eqMotion2}
and \eqref{eqMotion3} reduce to
 \begin{equation}
N_{1,\text{I}} = N_{-1,\text{I}} = 0,
\end{equation}  
which are to be satisfied for all positive $\lambda_1$, $\lambda_2$, 
$\lambda_3$. 
The corresponding constitutive equation is
 \begin{equation}\label{materialCB(-1)C=0}
\mathbb{T} = N_0(\text{I},\text{II},\text{III}) \mathbf{1} 
 + N_1(\text{II},\text{III}) \mathbb{B} 
  - N_{-1}(\text{II},\text{III}) \mathbb{B}^{-1}.
\end{equation}  
Again, the corresponding class of materials is inclusive of the
Generalized  Hadamard materials class.
When \eqref{CB(-1)C=0} holds, then 
 \begin{equation} 
\rho N^{-2}
  = (N_1 + 2 N_{0,\text{I}})\mathbf{C \cdot} \mathbb{B} \mathbf{C}.
\end{equation}  

To determine those $\mathbf{C}$ for which 
$\mathbf{C \cdot C} = \mathbf{C \cdot} \mathbb{B}^{-1} \mathbf{C}=0$,
we use the Hamiltonian decomposition of the $\mathbb{B}^{-1}$ tensor:
 \begin{equation} \label{hamiltonB(-1)}
\mathbb{B}^{-1} = \lambda_2^{-2} \mathbf{1}
 + \halft (\lambda_1^{-2} - \lambda_3^{-2})
   [\mathbf{l^+}\otimes\mathbf{l^-}+\mathbf{l^-}\otimes\mathbf{l^+}],
\end{equation}  
where
 \begin{align} 
&\mathbf{l^\pm} = \hat{\delta} \mathbf{i} \pm \hat{\phi} \mathbf{k},
\quad
\hat{\delta} = \sqrt{\frac{\lambda_2^{-2}-\lambda_1^{-2}}
                             {\lambda_3^{-2}-\lambda_1^{-2}}},
\quad
\hat{\phi} = \sqrt{\frac{\lambda_3^{-2}-\lambda_2^{-2}}
                             {\lambda_3^{-2}-\lambda_1^{-2}}},
\quad
\hat{\delta}^2 + \hat{\phi}^2 = 1,
\nonumber \\
&\hat{\delta}^2 \lambda_3^{-2} + \hat{\phi}^2 \lambda_1^{-2} 
    = \lambda_2^{-2},
\quad
\hat{\delta}^2 \lambda_3^2 + \hat{\phi}^2 \lambda_1^2
 -  \lambda_2^2 = \hat{\delta}^2\hat{\phi}^2 
       (\lambda_3^{-2} - \lambda_1^{-2})^2 \text{III}.
\end{align}  
Here the only possible $\mathbf{C}$ are those for which
$\mathbf{C \cdot C} = 0$, namely $\mathbf{C \cdot l^+} = 0$ or 
$\mathbf{C \cdot l^-} = 0$.
Thus, suitable $\mathbf{C}$ are
 \begin{equation} \label{CwhenCB(-1)C=0}
\mathbf{C} = \hat{\phi} \mathbf{i} \pm i \mathbf{j} 
                  - \hat{\delta} \mathbf{k},
\quad \text{or} \quad
\mathbf{C} = \hat{\phi} \mathbf{i} \pm i \mathbf{j}
                  + \hat{\delta} \mathbf{k}.
\end{equation}  
For all these four possible choices of $\mathbf{C}$, 
the corresponding complex scalar slowness $N$ is given by
 \begin{equation} \label{NwhenCB(-1)C=0}
\rho N^{-2} = 
  \hat{\delta}^2 \hat{\phi}^2
       (\lambda_3^{-2}-\lambda_1^{-2})^2 \text{III}
            (N_1 + 2 \text{III} N_{0,\text{I}}).
\end{equation}  

In order that the constitutive equation \eqref{materialCB(-1)C=0} 
represent that of a hyperelastic material, the response functions 
$N_0$, $N_1$, $N_{-1}$ must satisfy certain compatibility equations 
\cite[\S 86]{TrNo65}.
Using those compatibility equations it is found that for 
\eqref{materialCB(-1)C=0} the response function 
$N_1(\text{II}, \text{III})$ must be independent of $\text{II}$. 
Then integrating the compatibility equations it is found that  
\begin{equation}
N_1 = R(\text{III}), \quad 
N_0 = 
 \text{I } \text{III}^\halft \dfrac{\text{d}}{\text{dIII}}
     [R(\text{III})\text{III}^\halft] + S(\text{II}, \text{III}),
\end{equation}
where $R$  and $S$ are arbitrary functions of their arguments and 
$N_{-1}$ is a solution of 
\begin{equation}
\halft N_{-1} 
 + \text{III} \dfrac{\partial  N_{-1}}{\partial \text{III}} 
  + \text{II} \dfrac{\partial  N_{-1}}{\partial \text{II}}
   + \text{III}\dfrac{\partial  S(\text{I}, \text{III})}
                                         {\partial \text{II}} 
  = 0.
\end{equation}
The characteristics of this equation are
\begin{equation} \label{characteristics}
\dfrac{\text{dIII}}{\text{III}} =\dfrac{\text{dII}}{\text{II}} = 
\dfrac{\text{d}N_{-1}}
  {\halft N_{-1} + \text{III} \partial  S / \partial \text{II}},
\end{equation}
the solutions of which depend upon $S$. 
To make progress we assume 
\begin{equation} 
S = M(\text{III}) \text{II} + T(\text{III}), 
\end{equation} 
where $M$ and $T$ are arbitrary functions of $\text{III}$. 
In this case the general solution of  \eqref{characteristics} is 
\begin{equation} 
f(\text{II}/\text{III}, \text{ III}^\halft N_{-1}
 - \textstyle{\int} \text{III}^\halft M(\text{III}) \text{dIII}) = 0,
\end{equation}
where $f$ is an arbitrary function, or 
\begin{equation} 
N_{-1} = 
  - \text{III}^{-\halft} 
     \textstyle{\int} \text{III}^\halft M(\text{III}) \text{dIII} 
 + \text{III}^{-\halft} h(\text{II}/\text{III}), 
\end{equation}
where $h$ is an arbitrary function. 
It is found that $h$ must be zero for a hyperelastic material. 
We obtain 
\begin{equation} 
2 W = 
 \text{I} \; \text{III}^\halft R(\text{III}) 
  - \text{II }  \text{III}^{-1}
     \textstyle{\int} \text{III}^\halft M(\text{III}) \text{dIII} 
 + \textstyle{\int} \text{III}^{-\halft} T(\text{III}) \text{dIII}, 
\end{equation}
and 
\begin{align} 
& N_1 =  R(\text{III}), \quad 
 N_{-1} =  -\text{III}^{-\halft}
     \textstyle{\int} \text{III}^\halft M(\text{III}) \text{dIII}, 
\nonumber \\ 
&  N_0 =  \halft  \text{I} R(\text{III}) + \text{I III} R'(\text{III})
                + \text{II}M(\text{III}) +  T(\text{III}), 
\end{align}
where $R$, $M$, $T$ are arbitrary functions of $\text{III}$. 
Of course alternative choices of $S$ will lead to alternative forms 
for $W$. 

Finally, we recap for the constitutive model \eqref{materialCBC=0} 
(or \eqref{materialCB(-1)C=0}) that amongst the waves which may 
propagate in the material when it is held in an arbitrary state of 
finite static homogeneous deformation are two infinitesimal 
circularly polarized longitudinal plane waves whose circles 
of polarization lie in the planes of the central circular sections 
of the $\mathbb{B}$-ellipsoid (or the $\mathbb{B}^{-1}$-ellipsoid) 
corresponding to the finite static deformation.

\subsection{Circle of polarization lies in a plane of central
circular section of the $\mathbb{B}^n$-ellipsoid.} 
\label{central circular section of the B(n)-ellipsoid}

Here we determine the most general form of the response functions 
$N_0$, $N_1$, $N_{-1}$ such that the circle of polarization lies in a 
plane of central circular section of the $\mathbb{B}^n$-ellipsoid, 
$\mathbf{x \cdot} \mathbb{B}^n \mathbf{x} = 1$, where $\mathbb{B}$ 
is the left Cauchy-Green strain tensor corresponding to the finite 
static homogeneous deformation.
As in the previous cases ($n=\pm 1$), it is found that the response 
functions $N_1$, $N_{-1}$ depend just upon two invariants of 
$\mathbb{B}$. 
They take the forms $N_\Gamma
 = N_\Gamma(\text{III}, \text{tr } \mathbb{B}^n)$.
Up to a sign there are four directional bivectors $\mathbf{C}$ such 
that two infinitesimal circularly polarized longitudinal plane waves 
may propagate and whose circles of polarization lie in the planes of 
central circular section of the $\mathbb{B}^n$-ellipsoid.

Now we have
\begin{equation} \label{CB(n)C=0}
 \mathbf{C \cdot C} =  
  \mathbf{C \cdot} \mathbb{B}^n \mathbf{C} = 0,
 \end{equation}  
and equations \eqref{3rdEqn}, \eqref{eqMotion2}, \eqref{eqMotion3} 
become
\begin{equation} \label{NforCB(n)C=0}
  \rho N^{-2} = 
  (2 N_{0,\text{I}} + N_1) \mathbf{C \cdot} \mathbb{B} \mathbf{C} 
    + (-2\text{III} N_{0,\text{II}} 
       + N_{-1}) \mathbf{C \cdot} \mathbb{B}^{-1} \mathbf{C},    
\end{equation}
and
 \begin{align}  \label{condCB(n)C=0}
& N_{1,\text{I}}(\mathbf{C \cdot} \mathbb{B} \mathbf{C}) 
   = \text{III} N_{1,\text{II}} 
       (\mathbf{C \cdot} \mathbb{B}^{-1} \mathbf{C}), 
\nonumber \\ 
& N_{-1,\text{I}}(\mathbf{C \cdot} \mathbb{B} \mathbf{C}) 
   = \text{III} N_{-1,\text{II}} 
        (\mathbf{C \cdot} \mathbb{B}^{-1} \mathbf{C}). 
\end{align}  

Now it may be shown (Appendix) that when 
\eqref{CB(n)C=0} hold, then 
\begin{equation} \label{resultAppendix}
(\mathbf{C \cdot} \mathbb{B} \mathbf{C}) / 
  (\mathbf{C \cdot} \mathbb{B}^{-1} \mathbf{C})
 = \text{III} (\partial \text{tr} \mathbb{B}^n / \partial \text{II})
            / (\partial \text{tr} \mathbb{B}^n / \partial \text{I}), 
\end{equation}
so that \eqref{condCB(n)C=0} become 
\begin{equation} \label{condCB(n)C=0(2)}
N_{\Gamma, \text{I}} 
    \dfrac{\partial \text{tr} \mathbb{B}^n}{\partial \text{II}}
 - N_{\Gamma, \text{II}}
     \dfrac{\partial \text{tr} \mathbb{B}^n}{\partial \text{I}}
 = 0, 
\quad 
\Gamma = \pm 1. 
\end{equation}
This suggests that we write 
$N_\Gamma(\text{I}, \text{II}, \text{III})$ 
in terms of the set of invariants 
$\text{I}, \text{III}, \text{tr} \mathbb{B}^n$ ($n \ne 1$) which 
is equivalent to the set $(\text{I}, \text{II}, \text{III})$. 
We write 
\begin{equation} 
N_\Gamma = 
   \hat{N}_\Gamma(\text{I}, \text{III}, \text{tr} \mathbb{B}^n), 
\quad 
\Gamma = \pm 1,
 \end{equation}
and
\begin{align} \label{derivativesN}
& \dfrac{\partial N_\Gamma} {\partial \text{I}}
   =  \dfrac{\partial \hat{N}_\Gamma} {\partial \text{I}}
     + \dfrac{\partial \hat{N}_\Gamma} 
         {\partial \text{tr} \mathbb{B}^n}
       \dfrac{\partial \text{tr} \mathbb{B}^n}{\partial \text{I}}, 
\nonumber  \\
 & \dfrac{\partial N_\Gamma} {\partial \text{II}}
   =  \dfrac{\partial \hat{N}_\Gamma} {\partial \text{II}}
     + \dfrac{\partial \hat{N}_\Gamma} 
         {\partial \text{tr} \mathbb{B}^n}
      \dfrac{\partial \text{tr} \mathbb{B}^n}{\partial \text{II}},
\end{align}
so that \eqref{condCB(n)C=0(2)} becomes 
\begin{equation} 
\dfrac{\partial \hat{N}_\Gamma} {\partial \text{I}} = 0, 
\quad 
\Gamma = \pm 1. 
\end{equation}
and thus, because these are to be valid for all 
$\lambda_1, \lambda_2, \lambda_3 > 0$, we have 
\begin{equation} 
 \hat{N}_\Gamma = 
 \hat{N}_\Gamma (\text{III}, \text{tr} \mathbb{B}^n), 
\quad 
\Gamma = \pm 1. 
\end{equation}
Hence, those materials such that when they are in any state of 
finite static homogeneous deformation, two infinitesimal 
`longitudinal' inhomogeneous circularly polarized plane waves 
may propagate, the circle of polarization being in the planes 
of central circular sections of the $\mathbb{B}^n$-ellipsoid, 
where $\mathbb{B}$ is the strain tensor associated with 
the finite static homogeneous deformation, 
have constitutive equation 
\begin{equation} \label{constitutiveCB(n)C=0}
\mathbb{T} =  
 \hat{N}_0(\text{I}, \text{III}, \text{tr} \mathbb{B}^n)\mathbf{1} 
 + \hat{N}_1( \text{III}, \text{tr} \mathbb{B}^n)\mathbb{B} 
 - \hat{N}_{-1}(\text{III}, \text{tr} \mathbb{B}^n)\mathbb{B}^{-1}.
\end{equation} 
This result is in accord with what we found previously when 
$n = \pm 1$.
Indeed, for $n=1$, $\text{tr} \mathbb{B}^n = \text{I}$ and 
\eqref{constitutiveCB(n)C=0} becomes \eqref{materialCBC=0},
whilst for $n=-1$, $\text{tr} \mathbb{B}^n = \text{II}/\text{III}$ and 
\eqref{constitutiveCB(n)C=0} may be written in the form of 
\eqref{materialCB(-1)C=0}. 

Returning to the general case, upon using 
\eqref{NforCB(n)C=0},  \eqref{resultAppendix},  
and \eqref{derivativesN}, we find 
\begin{equation} \label{NforCB(n)C=0(2)}
\rho N^{-2} = 
  (2 \dfrac{\partial \hat{N}_0}{\partial \text{I}} + \hat{N}_1) 
         \mathbf{C \cdot} \mathbb{B} \mathbf{C} 
     +  \hat{N}_{-1} \mathbf{C \cdot} \mathbb{B}^{-1} \mathbf{C},
\end{equation} 

Thus, if $\mathbf{C}$ is chosen so that 
$ \mathbf{C \cdot C} =  
  \mathbf{C \cdot} \mathbb{B}^n \mathbf{C} = 0$, 
then a circularly polarized plane wave with slowness 
$\mathbf{S} = N\mathbf{C}$ ($N$ given by \eqref{NforCB(n)C=0(2)}) 
may propagate in the material with constitutive equation 
\eqref{constitutiveCB(n)C=0} when it is held in an arbitray state 
of finite static homogeneous deformation. 
The possible choices of $\mathbf{C}$ satisfying 
$ \mathbf{C \cdot C} =  
  \mathbf{C \cdot} \mathbb{B}^n \mathbf{C} = 0$ are 
\begin{equation}
\mathbf{C} = 
 \phi_n \mathbf{i} \pm  i\mathbf{j} - \delta_n \mathbf{k},
\quad 
\mathbf{C} = 
 \phi_n \mathbf{i} \pm  i\mathbf{j} + \delta_n \mathbf{k},
\end{equation}
where 
\begin{equation}
\delta_n = 
  \sqrt{\dfrac{\lambda_1^{2n} - \lambda_2^{2n}}
                              {\lambda_1^{2n} - \lambda_3^{2n}}},
\quad
\phi_n = 
 \sqrt{\dfrac{\lambda_2^{2n} - \lambda_3^{2n}}
                               {\lambda_1^{2n} - \lambda_3^{2n}}},
\quad
\delta_n^2 + \phi_n^2 = 1.
\end{equation}
We note that the terms occuring in \eqref{NforCB(n)C=0(2)}
for all four possible choices of $\mathbf{C}$ are given by 
\begin{align}
& \mathbf{C \cdot} \mathbb{B} \mathbf{C}
 = (\lambda_1^2 - \lambda_2^2)\phi_n^2 
       - (\lambda_2^2 - \lambda_3^2) \delta_n^2,
\nonumber \\  
& \mathbf{C \cdot} \mathbb{B}^{-1} \mathbf{C}
 = [\lambda_1^2(\lambda_2^2 - \lambda_3^2)\delta_n^2 
       - \lambda_3^2(\lambda_1^2 - \lambda_2^2) \phi_n^2]
    / \text{III}.
\end{align}

\subsubsection*{Special Case $n=\textstyle{\frac{1}{2}}$}

The expressions simplify greatly in the case when 
$n=\textstyle{\frac{1}{2}}$, that is, when we assume that the 
circles of polarization lie in the planes of central circular section 
of the $\mathbb{B}^{\textstyle{\frac{1}{2}}}$-ellipsoid.
Thus $\mathbf{C}$ is such that $\mathbf{C\cdot C}=0$, 
$\mathbf{C \cdot}\mathbb{B}^{\textstyle{\frac{1}{2}}}\mathbf{C} = 0$, 
so that the possible $\mathbf{C}$ are given by 
\begin{align}
& \sqrt{\lambda_1 - \lambda_3} \mathbf{C} =
 \sqrt{\lambda_2 - \lambda_3} \mathbf{i}
  \pm i\mathbf{j} + \sqrt{\lambda_1 - \lambda_2} \mathbf{k},
\nonumber \\
& \sqrt{\lambda_1 - \lambda_3} \mathbf{C} =
 \sqrt{\lambda_2 - \lambda_3} \mathbf{i}
  \pm i\mathbf{j} - \sqrt{\lambda_1 - \lambda_2} \mathbf{k}.
 \label{5.43}
\end{align}
The corresponding constitutive equation is 
\begin{equation} 
\mathbb{T} =  
 \hat{N}_0(\text{I}, \text{II}, \text{III} )\mathbf{1} 
 + \hat{N}_1( \text{III}, \text{tr}\mathbb{B}^\halft)\mathbb{B} 
 - \hat{N}_{-1}(\text{III}, \text{tr}\mathbb{B}^\halft)\mathbb{B}^{-1}.
\end{equation} 
Using \eqref{NforCB(n)C=0(2)}, the slownesses 
$\mathbf{S} = N \mathbf{C}$ of all four possible circularly polarized 
waves with directional bivectors $\mathbf{C}$ given by 
\eqref{5.43} are such that 
\begin{equation}
\rho N^{-2} = 
  (\lambda_1 - \lambda_2)(\lambda_2 - \lambda_3)
[2 \dfrac{\partial \hat{N}_0}{\partial \text{I}} + \hat{N}_1 
 + \dfrac{\lambda_1\lambda_2 + \lambda_2\lambda_3 + \lambda_3\lambda_1}
    {(\lambda_1 \lambda_2 \lambda_3)^2} \hat{N}_{-1}].
\end{equation}

\section*{Appendix} 
 \renewcommand{\thesection}{\Alph{section}}
\numberwithin{equation}{section}

Here we present a proof of \eqref{resultAppendix}. 

For convenience we write 
\begin{equation}
  \alpha = \lambda_1^2, \quad \beta = \lambda_2^2, \quad  
    \gamma = \lambda_3^2,  
\end{equation}
so that 
\begin{align}
& \mathbb{B} = \text{diag } (\alpha,  \beta,  \gamma), \quad
 \text{I} = \alpha + \beta + \gamma, \quad
 \text{III} = \alpha  \beta  \gamma, 
\nonumber \\
& \text{II} = \alpha \beta + \beta \gamma +\gamma \alpha, \quad
\text{tr} \mathbb{B}^n = \alpha^n + \beta^n + \gamma^n.
\end{align}

It may be checked that
\begin{equation}
  \dfrac{\partial \alpha}{\partial \text{I}} = 
\dfrac{\alpha (\beta - \gamma)\text{III}}
       {\beta\gamma(\alpha -\beta)(\beta - \gamma)(\gamma - \alpha)},
\quad 
 \dfrac{\partial \alpha}{\partial \text{II}} = 
\dfrac{ -(\beta - \gamma)\text{III}}
       {\beta\gamma(\alpha -\beta)(\beta - \gamma)(\gamma - \alpha)}, 
\end{equation}
and that
\begin{align}
&  \dfrac{\partial\text{tr} \mathbb{B}^n}{\partial \text{I}} = 
\dfrac{n[(\beta - \gamma)\alpha^{n+1}
                + (\gamma - \alpha)\beta^{n+1} 
                 + (\alpha -\beta)\gamma^{n+1}]}
       {(\alpha -\beta)(\beta - \gamma)(\gamma - \alpha)},
\nonumber \\
&  \dfrac{\partial\text{tr} \mathbb{B}^n}{\partial \text{II}} = 
\dfrac{-n[(\beta - \gamma)\alpha^n
                + (\gamma - \alpha)\beta^n 
                 + (\alpha -\beta)\gamma^{n}]}
       {(\alpha -\beta)(\beta - \gamma)(\gamma - \alpha)}.
\end{align}
Thus, we have the identity
\begin{equation}
  n \dfrac{\partial \text{tr} \mathbb{B}^{n+1}}{\partial \text{II}} 
 = - (n+1)
    \dfrac{\partial \text{tr} \mathbb{B}^n}{\partial \text{I}}. 
\end{equation}
If $\mathbf{C}$ is such that 
$\mathbf{C \cdot C} =  
  \mathbf{C \cdot} \mathbb{B}^n \mathbf{C} = 0$, i.e.
  \begin{equation}
    C_1^2 + C_2^2 + C_3^2 =0, \quad 
    C_1^2 \alpha^n + C_2^2 \beta^n + C_3^2 \gamma^{n} =0,
  \end{equation}
then 
\begin{equation}
  \dfrac{C_1^2}{C_3^2} = 
 \dfrac{\beta^n - \gamma^{n}}{\alpha^n - \beta^n}, \quad 
  \dfrac{C_2^2}{C_3^2} = 
 \dfrac{\gamma^n - \alpha^{n}}{\alpha^n - \beta^n},  
\end{equation}
so that 
\begin{align}
  \dfrac{\mathbf{C \cdot} \mathbb{B} \mathbf{C}}
            {\mathbf{C \cdot} \mathbb{B}^{-1} \mathbf{C}} & = 
  \dfrac{C_1^2\alpha + C_2^2\beta + C_3^2\gamma}
           {C_1^2/\alpha + C_2^2/\beta + C_3^2/\gamma} 
\nonumber \\
&  = 
 -\text{III} \dfrac{(\beta - \gamma)\alpha^n + 
                      (\gamma - \alpha) \beta^n + 
                        (\alpha - \beta)\gamma^n}
           {(\beta - \gamma)\alpha^{n+1} + 
                      (\gamma - \alpha) \beta^{n+1} + 
                        (\alpha - \beta)\gamma^{n+1}}.
\end{align}
Thus, when 
$\mathbf{C \cdot C} =  
  \mathbf{C \cdot} \mathbb{B}^n \mathbf{C} = 0$, we have the result 
  \begin{equation}
  \dfrac{\mathbf{C \cdot} \mathbb{B} \mathbf{C}}
            {\mathbf{C \cdot} \mathbb{B}^{-1} \mathbf{C}}=
 \text{III} 
   \dfrac{\partial \text{tr} \mathbb{B}^n /\partial \text{II}}
          {\partial \text{tr} \mathbb{B}^n /\partial \text{I}}. 
  \end{equation}



\end{document}